\documentclass[twocolumn,showpacs,preprintnumbers,amsmath,amssymb,prb]{revtex4}
\usepackage{graphicx}
\usepackage{dcolumn}
\usepackage{bm}

\begin{document}

\title{Collective Decoherence of Nuclear Spin Clusters}

\author{A. Fedorov}
\author{L. Fedichkin}

\affiliation{Center for Quantum Device Technology, Department of
Physics\\
and Department of Electrical and Computer Engineering,\\
Clarkson University, Potsdam, NY 13699--5721, USA}

\date{\today}

\begin{abstract}
The problem of dipole-dipole decoherence of nuclear spins  is
considered for strongly entangled spin cluster. Our results show
that its dynamics can be described as the decoherence due to
interaction with a composite bath consisting of fully correlated and
uncorrelated parts. The correlated term causes the slower decay of
coherence at larger times. The decoherence rate scales up as a
square root of the number of spins giving the linear scaling of the
resulting error. Our theory is consistent with recent experiment
reported in decoherence of correlated spin clusters.
\end{abstract}

\maketitle

\section{Introduction}
Quantum information processing devices are expected to be efficient
tool for solving some practical problems which are exponentially
hard for classical computers~\cite{Nielsen}. Their potential
computational performance is achieved by exploiting quantum
evolution of many particle system in exponentially large Hilbert
space, necessarily including evolution steps through entangled
states. Experimental implementation of Shor's quantum factoring
algorithm in seven spin-$1/2$ nuclei molecule have been
demonstrated~\cite{Vandersypen}.

The question of whether a scalable implementation of quantum
computer is possible in near future implies therefore the question
of whether one can protect the fragile entangled states from
destructive environment. The dynamics of coherence loss of entangled
many-particle clusters has attracted much attention recently. Some
authors simulated the noisy environment as a single bosonic bath
embracing whole cluster~\cite{Palma,Zanardi,Quiroga, Hilke}. An
alternative approach in which the noise sources acting on each
cluster constituent are uncorrelated was also
studied~\cite{Quiroga,additivity}. The realistic model of
environment will be somewhere between these two cases.  Still, the
quantitative account for partially correlated environment
complicates analysis much~\cite{Maierle}, even for two particle
system~\cite{Storcz}. Until recently experimental data  on
decoherence of large clusters of highly entangled particles were
 also unavailable. In 2004 the coherence dynamics of groups of up to 650
 entangled nuclear spins was observed for the first time~\cite{Krojanski}.
This paper is motivated by this experimental breakthrough indicating
the partial correlation of the environment.

In this paper, we derive the dependence of decoherence rate of large
spin clusters due to completely correlated and uncorrelated
perturbation. The results are generalized  to the system consisting
of nuclear spins $I=1/2$ experimentally studied in the
paper~\cite{Krojanski} by using solid-state NMR technique for
powdered adamantane samples. Our results show that its dynamics
resembles the decoherence due to interaction with a composite bath
 with a given ratio of correlated and uncorrelated terms. The
 dependence of decoherence rate on number of spins in the cluster
 was obtained.

This paper is organized as follows. The investigated system and
experimental procedures are described in Sec. \ref{system}. In Sec.
\ref{theory} we calculate the dynamics of NMR signal for the cases
of totally correlated/uncorrelated external perturbations and for
the experimental situation when decay is caused by internal
dipole-dipole interaction. Comparison with experiment data and
discussions are given in Sec. \ref{discussion}. Concluding remarks
are summarized in Sec. \ref{summary}.
\begin{figure}[b]
\includegraphics[width=8cm , height=3.5cm]
{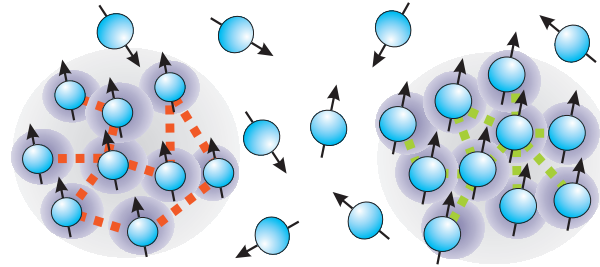} \caption{(Color online). Sketch of random array of
nuclear spins. Two clusters in partially correlated environments are
shown.} \label{sketch}
\end{figure}

\section{System}\label{system}

Our study was stimulated by the recent results of Krojanski and
Suter~\cite{Krojanski}. In their experiments the system of nuclear
spins-1/2 (protons) of the powdered admamantane sample was explored
by methods of NMR. Initially a system, placed in the external
magnetic field $H_0$ along $z$ axes, is in thermal equilibrium
\begin{equation}
\rho_{eq} = \frac{\textbf{1}}{2^N } + \frac{\gamma \hbar H_0}{kT}
\sum\limits_j {I_z^j },
\end{equation}
where $N$ is number of spins, $\gamma$ is spin gyromagnetic ratio,
$k$ is Boltzmann constant, $T$ is temperature and $I^{i}_{z}$ is $z$
component of $i$-th spin operator.
 With the help of special sequence of radio-frequency
pulses~\cite{Krojanski} the high-order correlations between spins
grow thereby creating an ensemble of weakly coupled spin clusters.
As a result, to describe evolution of spins in the sample it
suffices to consider only the dynamics of one such cluster with well
defined number of spins $n$~\cite{Krojanski,Baum, Lacelle}.

Existence of high-order coherences in $n$-spin system can be
formally described by presence of the off-diagonal elements
$\rho_{ij}$ of the spin density operator in any representation whose
basis states can be characterized by the total quantum magnetic
numbers: $M_i|i\rangle=I_z|i\rangle$, $M_j|j\rangle=I_z|j\rangle$.
Following the notation used in multiple quantum NMR
experiments~\cite{Slichter} we say that every off-diagonal density
matrix element $\rho_{ij}$ represents the coherence of the order $M$
where $M=M_i-M_j$. The number of coherences (different off-diagonal
elements) of the order $M$ in $n$-spin system at large $n$ is given
by
\begin{equation}\label{2}
C_{2n}^{n+M}=\frac{{(2n)!}}{{(n - M)!(n + M)!}}  \simeq
\frac{{2^{2n} }}{{\sqrt {\pi n} }}\exp \left( -
\frac{M^2}{n}\right).
\end{equation}
It is conventional to assume that after long pulse sequence spins
are prepared in the state described by the density operator
$\rho(0)$ with all even coherences excited with equal
probability~\cite{Baum, Krojanski}.

After the system is prepared in this high-correlated state it decays
under dipole-dipole interaction given by the Hamiltonian
\begin{equation}\label{3}
H_{dd}= \sum\limits_{j < k} d_{jk} \left( 3I_z^j I_z^k
-\bf{I}^j\cdot \bf{I}^k \right),
\end{equation}
where $d_{jk}  =  \frac{1}{2}\hbar ^2 \gamma ^2(1 - 3\cos ^2 [\theta
_{jk} ])/r_{jk}^3 $ and  $r_{jk}$, $\theta _{jk}$ are corresponding
absolute value and the angle with $z$ direction of the vector
connecting  $j$-th and $k$-th spins.

The system, evolving according to
\begin{equation}\label{}
\rho(t)=\exp\left(-\frac{i}{\hbar}H_{dd}t\right)\rho(0)\exp\left(\frac{i}{\hbar}H_{dd}t\right),
\end{equation}
does not produce experimentally observable signal. To analyze the
effect of dipole-dipole interaction, it undergoes conversion step by
another sequence of radio-frequency pulses described in
Ref.~\cite{Baum}. During this step multiple-quantum coherences are
converted back to single-quantum longitudinal magnetization. After
applying a resonant frequency $\pi/2$ pulse which converts the
longitudinal magnetization into transverse one the resulting
longitudinal magnetization can be determined by measuring the free
induction decay. The free induction decay amplitude right after
$\pi/2$ pulse is proportional to
\begin{equation}\label{4}
S(t) \propto {\rm{Tr}}\left[ {\rho (t)\rho (0)} \right],
\end{equation}
where $t$ is the time the system freely evolved under dipole-dipole
Hamiltonian between the end of the preparation step and the
beginning of the conversion step. The experiment has to be repeated
for sequence of decay times $t$ to obtain the decay of coherence.
The overall signal can be presented as a sum of contributions
corresponding to different coherence orders $M$~\cite{Krojanski}
\begin{equation}\label{8}
 S(t)=\sum_M S_M (t).
\end{equation}
The decay times for $S_M(t)$ were also measured
experimentally~\cite{Krojanski} as a function of coherence order $M$
for different cluster sizes $n$.

\section{Theory} \label{theory}
\subsection{Decay of NMR signal due to uncorrelated/correlated external baths}
First of all, consider a model when the decay of coherence occurs
due to interaction with the external bath. We do not specify the
bath itself and use the generic picture. In other words, in this
subsection we consider the system without dipole-dipole interaction
between spins. Instead we introduce some interaction with external
bath which causes the initial spin coherence to decay. In this paper
we focus only on the dephasing part of this interaction. A classical
analog of such model can be a cluster of spins in fluctuating
external magnetic field directed along $z$ axis~\cite{Maierle}.

Henceforward, we use Zeeman basis $|a\rangle=|a_1...a_n\rangle$,
where $a_i=\pm1$ and $I^i_z|a_i\rangle=(a_i/2)|a_i\rangle$. If we
consider the interaction of a single spin with the bath, its
evolution in Zeeman basis is given by
\begin{equation}\label{single}
    \rho_{\pm1,\pm1}(t)=\rho_{\pm1,\pm1}(0); \; \rho_{1,-1}(t)=\rho_{1,-1}(0)e^{-\Gamma(t)},
\end{equation}
where we used the interaction representation and the explicit form
of the decay function $\Gamma(t)$ is determined by the nature of
specific spin-bath interaction. Our main question is how the rate of
the collective decoherence of the correlated spin cluster, measured
by the technique given in the previous section, differs from the one
for the single spin dynamics (\ref{single}). The answer certainly
depends on degree of correlation of the bath at different spin
sites.

As the first example, we consider the limiting case  of completely
uncorrelated environment: each spin interacts with its own bath
assuming no correlations between baths related to different spins.
In this case the matrix elements of $n$ spin system density operator
evolve according to Ref.~\cite{Palma}
\begin{equation}\label{decay1}
  \rho_{ab}(t)=\rho_{ab}(0)\exp\left(-\Gamma_{ab}(t)\right),
\end{equation}
where collective decay function $\Gamma_{ab}(t)$ can be expressed in
term of single spin decay function $\Gamma(t)$ as
$\Gamma_{ab}(t)=f\Gamma(t)$ and $f=(1/2)\sum_i |a_i-b_i|$ is the
Hamming distance between the spin states $|a\rangle$ and
$|b\rangle$. The value of Hamming distance $f$ has the same parity
as coherence order $M$ and is within the limits $f\in[M,n]$. The
number of configurations for given $f$ and $M$ for the system of $n$
spin-1/2 can be found as
\begin{equation}\label{Cf}
    2^{n-f}C_n^f C_{f}^{\frac{f+M}{2}}\simeq\frac{2^{2n}}{\pi\sqrt {n f} }\exp \left[ -
\frac{(f-n/2)^2}{n/2}\right]\exp\left[-\frac{M^2}{2f}\right].
\end{equation}

We can calculate the observable decay of NMR signal $S(t)$ according
to (\ref{4},\ref{decay1}) as
\begin{equation}\label{S}
S(t) = \sum_{a,b}|\rho_{ab}(0)|^2\exp[-\Gamma_{ab}(t)],
\end{equation}
where we need to carry out the summation over all possible
amplitudes $|\rho_{ab}|$. The signal contributions $S_M(t)$ due to
certain coherence order $M$ can be evaluated by use of the same
formula (\ref{S}). Although in this case one needs to take the sum
over only the subset of configurations $\{|a\rangle\langle
b|\}\in\mathfrak{M}$ for whose the additional condition
$\sum\nolimits_j (a_j-b_j)=2 M$ is satisfied. The situation is
greatly simplified by assuming that all even coherences are
initially excited with equal probability: $|\rho_{ab}(0)|={\rm
const}$ if $(1/2)\sum\nolimits_j (a_j-b_j)=0,2,4,..$; while all
other coherences are not existent $\rho_{ab}(0)=0$~\cite{Krojanski,
Baum}. We can write
\begin{equation}\label{}
    S_M (t)\propto\sum_{a,b\subset\mathfrak{
M}}\exp\left(-if\Gamma(t)\right).
\end{equation}
Integrating over all $f$  with corresponding weight (\ref{Cf}) we
obtain, for $n\Gamma(t)\lesssim1$, $n\gg1$ and $M\leq n/2$,
\begin{equation}\label{n/2}
    S_M (t)=\exp\left(-\frac{n}{2}\Gamma(t)\right).
\end{equation}
We are interested in times up to $1/e$ decay time where formula
(\ref{n/2}) is valid. Moreover, since $n\Gamma(t)\lesssim1$ the
decay function is only in the onset regime: $\Gamma(t)\lesssim
1/n\ll 1$ for $n\gg1$. Therefore, we take only the lowest
non-vanishing order of decay function in time (which is always
quadratic)
\begin{equation}\label{Gt}
    \Gamma(t)=\alpha t^2+O(t^4).
\end{equation}
Using (\ref{n/2},\ref{Gt}) we obtain
\begin{equation}\label{n/2t2}
    S_M (t)=\exp\left(-\frac{n}{2}\alpha t^2\right).
\end{equation}
We emphasize that while we used the short-time expansion for the
decay function (\ref{Gt}) we expect the formula (\ref{n/2t2}) to be
valid up to $1/e$ decay time for the large number of spins $n\gg1$.

As a second example, we consider the case of completely correlated
environment when the whole cluster interacts with the same external
bath. The dynamics of the density matrix elements is given
by~\cite{Palma}
\begin{equation}\label{}
    \rho_{ab}(t)=\rho_{ab}(0)\exp\left(-M^2\Gamma(t)\right),
\end{equation}
and the signal $S_M(t)$ decays as
\begin{equation}\label{M2t2}
    S_M (t)=\exp\left(-M^2\Gamma(t)\right)\simeq\exp\left(-M^2\alpha t^2\right).
\end{equation}
Formula (\ref{n/2t2}) should be compared with (\ref{M2t2}). Both
results show that the decay of signal $S_M(t)$ can be approximated
by the Gaussian function up to $1/e$ decay times for $n\gg1$.
However, $M$ dependencies for two formulas are totally different.
The decay of $S_M(t)$ for uncorrelated environment does not
demonstrate any dependence on coherence order $M$ while for
correlated case it strongly depends on it. Thus, we established the
distinctive features of the influence of correlated/uncorrelated
environments onto spin cluster dynamics which can be observed
experimentally by NMR methods.

\subsection{Decay of NMR signal due to internal dipole-dipole interaction}
In the experiment by Krojanski and Suter~\cite{Krojanski} the
decoherence is caused not by external bath but due to integral
dipole-dipole interaction between spins. However, as we show below
the resulting behavior of the system can be interpreted with the
help of results obtained in the previous subsection.

The  dipole-dipole Hamiltonian (\ref{3}) commutes with Zeeman
Hamiltonian
\begin{equation}\label{Zeeman}
    H_{Z}=-\gamma \hbar H_0\sum_{j} I_z^j.
\end{equation}
However, the complexity of the system and especially the fact that
two terms $I_z^j I_z^k$ and ${\bf I}^j\cdot {\bf I}^k$ do not
commute makes impossible to find the exact (analytical or numerical)
solution of the problem~\cite{Abragam, Baum,Slichter}. Existence of
high order coherences in the state described by prepared density
operator $\rho(0)$ also complicates the application of the
traditional method of moments which enables to describe the decay of
coherence without solving explicitly for eigenvalues and eigenstates
of energy in case of single-quantum NMR experiments
\cite{VanVleck,Slichter}. In our case the decay of signal is not
proportional to the autocorrelation function ${\rm Tr} \{I_x(t)
I_x\}$, as in the case of decay of free induction signal
\cite{Abragam}, but is given by density operator correlator
(\ref{4}) where one needs to evaluate the summation of exponentially
large number of terms. In order to obtain the analytical results, we
focus on pure dephasing effect of dipole-dipole interaction
neglecting any spin exchange between spins that is described by
flip-flop term $I_x^j I_x^k +I_y^j I_y^k $ in Hamiltonian (\ref{3}).
The dipole-dipole dephasing Hamiltonian has the form
\begin{equation}\label{5} H_{dd}^{*}=
2\sum\limits_{j < k} d_{jk} I_z^j I_z^k.
\end{equation}
Because dephasing is not associated with energy transfer mechanism
it is generally the fastest source of decoherence~\cite{Mozyrski,
Hu}. It becomes the sole process for decoherence in the limit of
"unlike spins" \cite{Abragam,Slichter} when spin exchange is
suppressed. The consideration of only this type of
interaction enables analytical calculations which are also justified
by good agreement with experiment in wide
range of parameters as it will be demonstrated below.

\begin{figure}[t]
\includegraphics[width=9.5cm, height=7.5cm]
{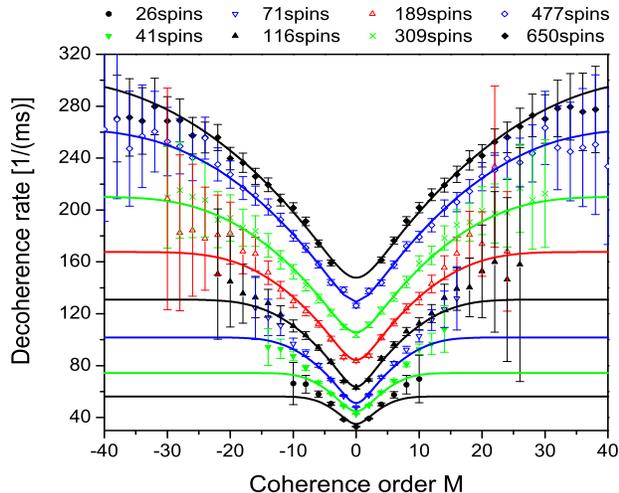} \caption{(Color online). Decoherence rate as function of
coherence order for different spin cluster sizes. The points
represent experimental values~\cite{Krojanski}. The solid lines are
obtained with accordance to theoretical formula (\ref{S_M}).  Degree
of correlation $p$ and Van Vleck second moment evaluated from
comparison with experiment are given in Table 1.} \label{decay}
\end{figure}

\begin{table}[b]
\caption{\label{tab:table2} Degree of correlation and second moment
for C$_{10}$H$_{16}$ obtained from decoherence rates for different
cluster sizes. }
\begin{ruledtabular}
\begin{tabular}{ccccccccc}
$n$ &26&41&71&116&189&309&477&650\\
\hline \\  $M_2$, $10^9 $s$^{-2}$ &1.50&1.65&1.60&1.60&1.65&1.60&1.60&1.55\\
\hline
p&0.27& 0.28& 0.33& 0.33& 0.32& 0.32& 0.32& 0.32\\
\end{tabular}
\end{ruledtabular}
\end{table}

In the Zeeman representation the off-diagonal density matrix
elements evolve according to (\ref{decay1}) where decay function is
given by
\begin{equation}\label{}
    \Gamma_{ab}(t)=
    \frac{it}{2}\sum_{j<k}d_{jk}(a_ja_k-b_j
    b_k).
\end{equation}

 The dynamics of normalized NMR signal (\ref{S}) can be
analytically expressed as
\begin{equation}\label{7}
S(t)=16\prod\limits_{j < k} \sum\limits_{ a_l,b_l=\pm 1; \atop
     l\neq k,j} \left| {\rho_{ab} (0)} \right|^2 {\cos ^2
(\frac{1}{2}d_{kj} t)}.
\end{equation}
Exact analytical expression (\ref{7}) does not provides us with much
information yet. Specifically, we intend to obtain explicit
dependence on number of spins in the cluster. For this purpose we
again assume that all even coherences are initially excited with
equal probability and the size of the cluster is large
$n\gg1$~\cite{Krojanski,Baum}. After performing some algebra (the
details are given in Appendix) we obtain the expression for
normalized signal
\begin{equation}\label{S_M1}
 S_M(t) =1-pM^2\frac{\alpha t^2}{2} -(1-p)\frac{n}{2}\frac{\alpha t^2}{2} +O(t^4),
\end{equation}
 in the second order in time. Here $ \alpha= M_2/9$ where $M_2=(9/4)\hbar^{-2}\sum_j d^2_{jk}$ is Van Vleck expression
for the second moment~\cite{Abragam} and degree of correlation $p$
is defined as
\begin{equation}\label{10}
  p=\frac{1}{n}\left(\sum\nolimits_j d_{jk}\right)^2\left/\sum\nolimits_j
  d^2_{jk}\right. ,
\end{equation}
so that $0\leq p \leq 1$. Formula (\ref{S_M1}) is valid only at
short time scales $n \alpha t^2\ll1$, while we are also interested
in much larger times up to $n \alpha t^2\sim1$. However, expansion
of the signal $S_M(t)$ in higher orders in time becomes exceedingly
difficult. Therefore, to continue (\ref{S_M1}) to the longer times
we use the analogy with the investigated limiting cases
(\ref{n/2t2}, \ref{M2t2}). Formula (\ref{S_M1}) contains two terms
proportional to $M^2$ and $n/2$ which can be regarded as
contributions from correlated and uncorrelated perturbations to spin
dynamics, respectively. In fact, the interaction described by
Hamiltonian (\ref{5}) can be semiclassically interpreted as the
perturbing magnetic field at the site of each spin (parallel or
antiparallel to the strong external magnetic field) produced by all
other spins in a cluster. The consequent spread of Larmor
frequencies for different spins in the cluster causes destructive
interference, or dephasing, observable by the decay of NMR signal.
The limit of totally correlated perturbation $p=1$ corresponds to
the case $d_{jk}\equiv\rm{const}$ leading to the same perturbing
field for each spin in the cluster. In contrast, the case of
absolutely random coefficients $\langle d_{jk}\rangle_j=0$ gives
$p=0$ and fully uncorrelated dynamics. The realistic situation is
expected to be in between these two limiting cases. Thus, we write
(\ref{S_M1}) as
\begin{equation}\label{S_M}
 S_M(t)  = p\exp\left(-M^2  \alpha t^2\right) + (1 -
 p)\exp\left(-\frac{n}{2} \alpha t^2\right),
\end{equation}
which is mathematically exact in up to the second order in time but
continued to the longer times $n \alpha t^2\sim1$. The total
magnetic resonance signal from the cluster $S(t)$ can be obtained by
summation over all contributions from different coherence orders
$S_M(t)$ according to formulas (\ref{8}) and (\ref{S_M})
\begin{equation}\label{S(t)}
 S(t)  = \frac{p}{\sqrt{n \alpha t^2+1}} + (1 -
 p)\exp\left(-\frac{n}{2} \alpha t^2\right).
\end{equation}
In order to understand whether the obtained formulas (\ref{S_M},
\ref{S(t)}) adequately describe the real experimental situation we
should check them with experiment data. The comparison of presented
theory and experiment is given in the next section.

\section{Comparison of theory with experiment and Discussions}\label{discussion}
Recent experiments~\cite{Krojanski} allowed us to estimate the
degree of correlation parameter for spin clusters in adamantane
samples. In Fig.~\ref{decay} we show curves of decay rates of
various coherence orders for different cluster sizes fitted to
experimental points. The decoherence rate was defined as the inverse
of $1/e$ decay time and was evaluated by solving the algebraic
equation $S_M(t)=1/e$ where $S_M(t)$ is given by formula
(\ref{S_M}). The degree of correlation $p$ and Van Vleck second
moment $M_2$ were extracted with the use of MATLAB software by
weighted least squares fitting to experimental data for every
cluster size $n$. We minimized $\sum_i (f(x_i)-f_i)^2/\Delta_i^2$
where $x_i$ and $f_i$ are experimental points, $f(x_i)$ are
corresponding theoretical solutions and $\Delta_i$ are experimental
errors denoted by vertical bars in Fig.~\ref{decay}. Obtained values
of $p$ and $M_2$ are given in Table 1. As it follows from formula
the definition of second moment~\cite{VanVleck} it is determined by
geometrical configurations and do not depend on cluster size $n$.
Its moderate fluctuations around average value ($\overline
M_2=(1.60\pm0.05)\cdot 10^9 {\rm s^{-2}}$) can be attributed to
experimental errors and corrections at small $n$. Obtained values
for the second moment are comparable but not identical with the
previous theoretical estimates and experimental measured values
$M_2\simeq2.6\cdot 10^9 {\rm s^{-2}}$ for powdered solid
adamantane~\cite{McCall,Smith}. The difference can be explained
either by crudity of the chosen model and neglecting flip-flop terms
or by discrepancy in the adamantane samples used in different
experiments. The question could be resolved by additional
measurement of the second moment $M_2$ for the given sample.

\begin{figure}[t]
\includegraphics[width=9.5cm, height=8.5cm]
{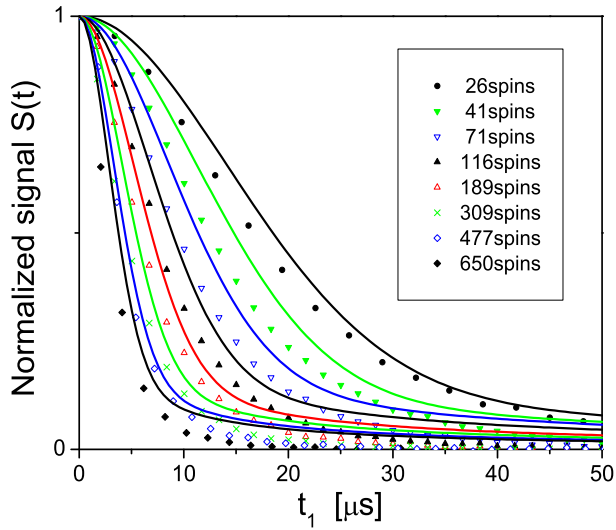} \caption{(Color online). Decay of coherence from
high-correlated spin clusters for different spin cluster sizes. The
points represent experimental values~\cite{Krojanski}. The solid
lines are values predicted by formula (\ref{S(t)}). } \label{rate}
\end{figure}

Taking the average value of $M_2=1.6\cdot 10^9 {\rm s^{-2}}$ and
values for $p$ from Table 1 it is possible to predict the temporal
dependence of total NMR signal from high-correlated spin cluster
(\ref{S(t)}) which was measured independently~\cite{Krojanski} for
different cluster sizes. The results shown in Fig.~\ref{rate} are in
good agreement with experiment. As can be seen from Fig.~\ref{rate}
the formula (\ref{S(t)}) describes the initial fast drop of
coherence with reasonable accuracy.The divergence at large times
between formula (\ref{S(t)}) (exact up second order in time) and
experimental results can be attributed by the contribution of higher
order terms.

\begin{figure}%[b]
\includegraphics[width=9.5cm, height=8.5cm]
{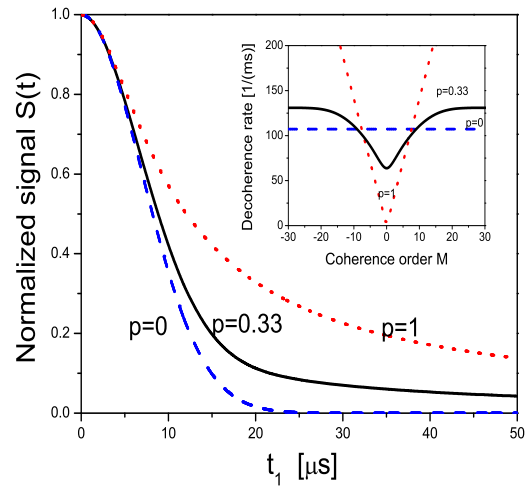} \caption{(Color online). Examples of temporal dependence
of the signal from high correlated spin cluster with size n=116 and
three values of degree of correlation $p$ for perturbation: p=0
(dashed line, uncorrelated perturbation), 0.33 (solid line, partial
correlation corresponding to experimental situation), 1 (dotted
line, correlated perturbation). The inset shows the decoherence rate
as a function of coherence order $M$.}\label{examples}
\end{figure}

Let us note, that the measured values of decoherence rates for
$S_M(t)$ give us only one time point in the temporal dynamics of the
signal for each value of $M$ and $n$. The parameters $M_2$ and $p$
obtained from fitting the solution of $S_M(t)=1/e$ to the
experimental data allow us to reconstruct the total decay dynamics
of $S_M(t)$ up to $1/e$ decay times. The following integration over
all $M$ provides the decay of overall signal $S(t)$ which was
measured independently. This procedure does not, by any means,
automatically guarantee the agreement of calculated values of $S(t)$
with experimentally measures ones. The correspondence of theoretical
to experimental values demonstrates the good degree of consistency
of presented theory.

Formula (\ref{S(t)}) allows us to analyze the influence of degree of
correlation on spin dynamics. Fig.~\ref{examples} shows the decay of
NMR signal for the spin cluster size of intermediate size $n=116$
and three representative examples of degree of correlation $p$:
$p=0$ (uncorrelated dynamics), $p=0.33$ (partially correlated
dynamics corresponding to the experimental situation) and $p=1$
(correlated dynamics). One can see that initially all three curves
decay equally. However, at later times the signal from the spin
cluster subject to correlated perturbation exhibits slower decay
compared to uncorrelated perturbation. That result comes from the
behavior of decoherence rate as function of coherence order $M$. As
can be seen from inset of Fig.~\ref{examples}, for uncorrelated
perturbation all coherence orders decay with the same, comparatively
high, rate $(n\alpha/2)^{1/2}$. In contrast, the decay rate for
correlated spin dynamics increases linearly  with absolute value of
$M$ as $\alpha^{1/2}|M|$. For the most probable configurations,
which according to (\ref{2}) are those with $M\approx0$, the decay
rate for correlated perturbation is actually less than that for
uncorrelated perturbation. The fact that correlated environment is
acting more delicate on specific groups of states is not surprising.
In particular, quantum computing error avoiding schemes based on
decoherence free subspaces~\cite{Zanardi, Lidar} are based on this
property.

For implementation of large-scale quantum computation the scaling of
decoherence rate with number of qubits is important. From the
expression (\ref{S(t)}) it transpires that decoherence rate
 of a spin cluster defined as inverse $1/e$ decay time \emph{always} increases as
$\propto \sqrt{n}$ with number of spins $n$ although the
corresponding factor depends on degree of correlation $p$. The
square root of $n$ scaling was indeed experimentally discovered
recently by Krojanski and Suter~\cite{Krojanski}.

For quantum information processing applications it is also important
to evaluate the error of a quantum computer, represented by a
cluster of high correlated spins, induced by dipole-dipole
interaction between spins. The error is defined as deviation of NMR
signal from its initial value due to decoherence processes during
the time required for elementary gate operation $t_g$:
$\delta_n=1-S(t_g)$. In order to provide successful implementation
of quantum error correction schemes, one needs to maintain this
error below the small threshold guarantying fault-tolerance
operation of these procedures~\cite{Nielsen}. Taking the smallness
of the parameter $\delta_n$ into account one can use (\ref{S(t)}) to
obtain
\begin{equation}\label{error}
   \delta_n\propto n t^2.
\end{equation}
This shows that if the error is small it scales linearly with number
of spins independently of degree of correlation. The linear scaling
of error agrees with theoretical results for bosonic models of
environment~\cite{additivity,Hilke} and suggests that the worst case
scenario of "superdecoherence"~\cite{Palma} is not realized for this
particular system.

\section{Summary} \label{summary}
In summary, we have presented a theory of coherence decay of
entangled spin clusters states due to internal dipole-dipole
interactions. Its dynamics resembles the decoherence due to
interaction with a composite bath consisting of fully correlated and
uncorrelated parts. The perturbation due to correlated terms leads
to the slower decay of coherence at larger times. The decoherence
rate scales up as a square root of the number of spins giving the
linear scaling of the resulting error. The results obtained can be
useful in analysis of decoherence effects in spin-based quantum
computers.

\begin{acknowledgments}
We are grateful to V.~Privman for suggesting the topic of this
research, and to H.~G.~Krojanski, D.~Solenov and D.~Suter for helpful
communications. This research was supported by the National Science
Foundation, Grant DMR-0121146, and by the National Security Agency
and Advanced Research and Development Activity under Army Research
Office Contract DAAD 19-02-1-0035.
\end{acknowledgments}

\appendix
\section{sdf}
The signal contributions $S_M(t)$ can be evaluated by use of the
formula (\ref{S})
\begin{equation}\label{A1}
S_M(t) =\sum_{a,b\subset\mathfrak{
M}}|\rho_{ab}(0)|^2\exp[-\Gamma_{ab}(t)]
\end{equation}
Here domain $\mathfrak{ M}$ denote all configurations
$\left|a\rangle\langle b \right|=\left|a_1\ldots a_n\rangle\langle
b_1 \ldots b_n \right|$ related to the certain coherence order $M$:
\begin{equation}\label{condition}
 \sum_j (a_j-b_j)=2 M.
\end{equation}
For every configuration $\left|a\rangle\langle b \right|$  we can
divide the total set of $n$ spins into two subsets $\mathbb{E}$ and
$\mathbb{N}$,
\begin{equation}\label{A2}
a_i=b_i, \;\forall i\in \mathbb{E};\quad a_j=-b_j,\;\forall
j\in\mathbb{N}.
\end{equation}
By the use of definitions (\ref{A2}) the decay function
$\Gamma_{ab}(t)$ can be simplified as
\begin{widetext}
\begin{eqnarray}\label{A3}
\Gamma_{ab}(t)&=&i\frac{t}{4}\sum_{j,k}d_{jk}(a_ja_k-b_j b_k)=
i\frac{t}{4}\sum_{j\in\mathbb{E},k\in\mathbb{E}}d_{jk}(a_ja_k-b_j
b_k)+
i\frac{t}{4}\sum_{j\in\mathbb{N},k\in\mathbb{N}}d_{jk}(a_ja_k-b_j
b_k)\nonumber\\
&+&i\frac{t}{2}\sum_{j\in\mathbb{E},k\in\mathbb{N}}d_{jk}(a_ja_k-b_jb_k)=it\sum_{j\in\mathbb{E},k\in\mathbb{N}}d_{jk}a_ja_k.
\end{eqnarray}
\end{widetext}

Assuming that all even coherences are initially excited with equal
probability, namely $|\rho_{ab}(0)|={\rm const}$ for all
$a,b\in\mathfrak{ M}$ (and for all other even order coherences), we
obtain the following formula for the decay of $S_M(t)$ according to
(\ref{A1},\ref{A3})
\begin{equation}\label{A4}
    S_M (t)\propto\sum_{a,b\subset\mathfrak{
M}}\exp[-it\sum_{j\in\mathbb{E},k\in\mathbb{N}}d_{jk}a_ja_k].
\end{equation}
We can redistribute the summation in (\ref{A4}) in the following way
\begin{equation}\label{}
   \sum_{a,b\subset\mathfrak{M}}=\sum_{\mathbb{E},\mathbb{N}}\sum_{a_j,a_k\atop i\in
   \mathbb{E},k\in\mathbb{N}}.
\end{equation}
Here the first sum in the left part of the equation is over all
possible choices of subsets $\mathbb{E}, \mathbb{N}$ in the set of
$n$ spins and second sum is over all possible values of $a_j,a_k$
for $i\in \mathbb{E},k\in\mathbb{N}$.
 It is easy to see that values $a_j$ for
$j\in\mathbb{E}$ can take any values $a_j=\pm 1$ since they do not
contribute to (\ref{condition}) and, therefore, do not change
coherence order $M$. We still have the condition for the values
$a_k$, $k\in\mathbb{N}$:
\begin{equation}\label{A5}
    \sum_{k\in\mathbb{N} } a_k =M.
\end{equation}
Thus, we can evaluate the summation over $a_j$ for $j\in\mathbb{E}$
first and obtain
\begin{eqnarray}\label{}
    S_M (t)&\propto& \sum_{\mathbb{E},\mathbb{N}}\sum_{a_k, k\in\mathbb{N}}\prod_{j\in\mathbb{E}}
    \cos(t\sum_{k\in\mathbb{N}}d_{jk}a_k)\\
    &\propto& \sum_{\mathbb{E},\mathbb{N}}\sum_{a_k,
    k\in\mathbb{N}}\left(1-\frac{t^2}{2}\sum_{j\in\mathbb{E}}(\sum_{k\in\mathbb{N}}d_{jk}a_k)^2+O(t^4)\right).\nonumber
\end{eqnarray}
Now we consider the term
$\sum\limits_{j\in\mathbb{E}}(\sum\limits_{k\in\mathbb{N}}d_{jk}a_k)^2$
to express it in terms of parameters of the material. We write
\begin{equation}\label{A6}
    d_{jk}=\bar d_j+\delta_{jk},
\end{equation}
where average coupling constant is defined as
\begin{equation}\label{}
\bar d_j=f^{-1}\sum_{k\in\mathbb{N}}d_{jk}.
\end{equation}
Here $f$ is Hamming distance between $|a\rangle$ and $|b\rangle$ or
the number of spins in subset $\mathbb{N}$. Note, that
$\sum_{k\in\mathbb{N}}\delta_{jk}=0$. By use of (\ref{A5},\ref{A6})
we obtain
\begin{equation}\label{}
(\sum_{k\in\mathbb{N}}d_{jk}a_k)^2= (\bar
d_j)^2M^2+\sum_{k\in\mathbb{N}}\delta_{jk}^2,
\end{equation}
where we neglected cross-terms $\sum_{k\in\mathbb{N}}\delta_{jk}a_k$
whose contribution is negligible for the large cluster sizes. For
$n\gg1$ we also approximate the summation over subsets by the
summation over total cluster with correction to the number of terms
in the sum:
$\sum_{i}=\left(n/(n-f)\right)\sum_{j\in\mathbb{E}}=(n/f)\sum_{j\in\mathbb{N}}$.
These assumptions should lead to the asymptotically correct value of
$S_M(t)$ for $n\rightarrow\infty$. We then evaluate the the signal
decay as
\begin{equation}\label{AS_M}
 S_M(t) \propto \sum_{\mathbb{E},\mathbb{N}} \left[1-\frac{t^2}{2}\frac{n-f}{n}\left(pM^2+(1-p)f\right)\sum_i
 d_{ki}^2\right]+O(t^4),
\end{equation}
where the parameter $p$ is defined as
\begin{equation}\label{Ap}
  p=\frac{1}{n}\left(\sum\nolimits_j d_{jk} \right)^2/\sum\nolimits_j
  d^2_{jk}.
\end{equation}
 After integration over all possible $f$ we deduce the closed, analytical form for the signal, exact up
to second order in time,
\begin{equation}\label{AS_M1}
 S_M(t) =1-\frac{t^2}{2}\frac{M_2}{9}\left(pM^2+(1-p)\frac{n}{2}\right)+O(t^4),
\end{equation}
Here $M_2=(9/4)\hbar^{-2}\sum_j d^2_{jk}$ is Van Vleck expression
for the second moment~\cite{Abragam}.

\end{document}